\documentstyle[11pt]{article}
\begin{document}
\date{} 
\begin{flushright}
{MIT-CTP-3076}\\
{YITP-01-1}
\end{flushright}
\vspace{2cm}
\begin{center}
{\Large {\bf Bulk Gauge Fields in the Bigravity Model}
\vskip0.8truein
{\large Motoi Tachibana}~\footnote{E-mail:{\tt motoi@lns.mit.edu}
~~JSPS Research Fellow\\
on leave from Yukawa Institute for Theoretical Physics,
Kyoto University, Kyoto 606-8502, Japan}\\}
\vskip0.4truein
\centerline{ {\it 
Center for Theoretical Physics,
Massachusetts Institute of Technology, Cambridge, MA 02139, USA.}}
\end{center}
\vskip0.8truein
\centerline{\bf Abstract}
\vskip0.3truein
Motivated by a recently proposed \lq \lq bigravity" model with two 
positive tension $AdS_4$ branes in $AdS_5$ by Kogan et al.[hep-th/0011141], 
we study behavior of bulk gauge field in
the model. In this case,
the zero mode of the gauge field is not constant but depends
on the fifth dimensional coordinate transverse to the brane. 
The zero mode then becomes
massive on the brane. From the physical requirement that the mass
must be small, a parameter of the model is constrained.
We also discuss the Kaluza-Klein modes.
\vskip0.13truein
\baselineskip=0.5 truein plus 2pt minus 1pt
\baselineskip=18pt
\addtolength{\parindent}{2pt}
\newpage
Over the past few years, theories with extra spatial dimensions
have received much attention because they potentially
solve long standing problems such as the hierarchy problem,
as was originally suggested by Antoniadis,
Arkani-Hamed, Dimopoulos and Dvali\cite{add}.
Afterwards Randall and Sundrum found a new solution to the
hierarchy problem\cite{rs1}. In addition, they showed that gravity 
could be localized on a brane with infinite extra dimension\cite{rs2}.
Since then, various types of models including gravity have been studied.

Recently Kogan et al. proposed a new \lq \lq bigravity" 
model\cite{kogan}. The model
has only two positive tension $AdS_4$ branes in $AdS_5$ bulk and no 
negative tension branes. Owing to the absence of negative tension branes,
the model does not have the ghost fields which appeared 
in the previous \lq \lq bigravity"
scenario\cite{kmprs} and also in the quasi-localized gravity
model\cite{grs}. But interestingly in the model of \cite{kogan}
the bounce of the warp factor mimics the effect of a negative 
tension brane and thus gives rise to an
anomalously light graviton Kaluza-Klein mode. Moreover it is possible
in this model to circumvent 
the van-Dam-Veltman-Zakharov no go theorem\cite{dvz}
about the non-decoupling of the extra polarization states of the massive
graviton\cite{kmp2}. Similar discussions have recently appeared in \cite{kr}.

In this paper, we study behavior of the bulk gauge field in this
new \lq \lq bigravity" model. We are especially interested in how
the zero mode of the gauge field behaves in this background.

Following the work by Kogan et al.\cite{kogan}
we shall consider a five dimensional
anti-de Sitter spacetime ($AdS_5$) with a warp factor
\begin{equation}
ds^2 = \Omega^2(w)(\bar{g}_{\mu \nu}(x)dx^{\mu} dx^{\nu} + dw^2),
\label{metric}
\end{equation}
where $\mu, \nu = 0,1,2,3.$ 
The metric $\bar{g}_{\mu \nu}(x)$
denotes an $AdS_4$ background. The warp factor $\Omega(w)$ in the
conformal coordinate $w$ is given by
\begin{equation}
\Omega(w)= \frac{1}{\cosh(kz_0)}\frac{1}{\cos\tilde{k}(|w|-\theta)},
\label{warp}
\end{equation}
where $\tilde{k} \equiv k/\cosh(kz_0)$ and
$\tan(\tilde{k}\theta/2)=\tanh(kz_0/2)$, while $\tanh(kz_0)
\equiv kV_1/|\Lambda|$. $k$ is the curvature of $AdS_5$ defined
through $k \equiv \sqrt{\frac{-\Lambda}{24M^3}}$. $\Lambda$ is
the five dimensional cosmological constant
which is negative. $V_1$ and $V_2$
are tensions of the 3-branes at the orbifold fixed points,
$w=0$ and $w=w_L$, respectively.

We are interested in behavior of the bulk gauge field in this background
metric. Let us start with the following five dimensional action;
\begin{equation}
S_{GF}=-\frac{1}{4}\int dx^4\int_{-w_L}^{w_L}dw
\sqrt{-G}G^{MN}G^{RS}F_{MR}F_{NS},
\label{action}
\end{equation}
where $M, N, R, S= 0,1,2,3,w$ and
$F_{MN}=\partial_M A_N -\partial_N A_M$. $A_M(x^{\mu}, w)$
is the bulk U(1) gauge field. The extension to non-Abelian gauge
field is straightforward. $G_{MN}$ is the five dimensional metric
defined through eq.(\ref{metric}).
The equations of motions are given by
\begin{equation}
\partial_M(\sqrt{-G}G^{MN}G^{RS}F_{NR})=0.
\label{eom}
\end{equation}
We solve the equations with the gauge condition 
\begin{equation}
A_w=0.
\label{gauge}
\end{equation}
As usual, we expand the field $A_{\mu}(x^{\mu}, w)$ into zero mode
and Kaluza-Klein modes:
\begin{equation}
A_{\mu}(x, w)=\sum_{n=0}^{\infty}a^{(n)}_{\mu}(x)\rho^{(n)}(w).
\label{kk}
\end{equation}
Plugging eq.(\ref{metric}) into eq.(\ref{eom}) with 
the gauge fixing condition (\ref{gauge}) and the mode expansion
(\ref{kk}), we have the following equations;
\begin{eqnarray}
\frac{1}{\sqrt{-\bar{g}}\bar{g}^{\nu \sigma}a_{\nu}^{(n)}}
\partial_{\mu}(\sqrt{-\bar{g}}\bar{g}^{\mu \rho}
\bar{g}^{\nu \sigma}f_{\rho \nu}^{(n)})
&=& -\frac{1}{\Omega\rho^{(n)}}\partial_w 
(\Omega\partial_w \rho^{(n)}), \nonumber \\
\Omega\partial_w \rho^{(n)} \partial_{\mu}
(\sqrt{-\bar{g}}\bar{g}^{\mu \nu}a_{\nu}^{(n)})&=&0.
\label{eom2}
\end{eqnarray}
Here $f_{\mu \nu}^{(n)}=
\partial_{\mu}a_{\nu}^{(n)}-\partial_{\nu}a_{\mu}^{(n)}$.

Both sides of the first equation in eq.(\ref{eom2})
must be constant, which we write as $m_n^2$. In addition, we
choose the Lorentz gauge in curved spacetime. Thus 
\begin{eqnarray}
\frac{1}{\sqrt{-\bar{g}}}
\partial_{\mu}(\sqrt{-\bar{g}}\bar{g}^{\mu \rho}
\bar{g}^{\nu \sigma}f_{\rho \nu}^{(n)})
&=& m_n^2 \bar{g}^{\nu \sigma}a_{\nu}^{(n)},  \nonumber \\
\partial_{\mu}(\sqrt{-\bar{g}}\bar{g}^{\mu \nu}a_{\nu}^{(n)})
&=&0.
\label{conditions}
\end{eqnarray}
Then the equation of motion for the bulk gauge field
reduces to
\begin{equation}
\partial_w (\Omega(w)\partial_w \rho^{(n)})+m_n^2\Omega(w)\rho^{(n)}=0.
\label{eom3}
\end{equation}
The zero mode of the gauge field ($m_n=0$) satisfies
\begin{equation}
\partial_w (\Omega(w)\partial_w \rho^{(0)}(w))=0.
\label{zmeq}
\end{equation}
Remarkably in the present case,
unlike the case of the Randall-Sundrum model\cite{rs2}, the zero mode is 
not constant but depends on the fifth dimensional coordinate $w$
\footnote{For recent work for the bulk gauge fields in the RS model,
see Refs.\cite{bgf}.}. 
It is of the form

\begin{equation}
\rho^{(0)}(w)=\frac{\alpha}{\tilde{k}}
\cosh(kz_0)\sin\tilde{k}(w-\theta) + \beta,
\label{zmsol}
\end{equation}
where $\alpha$ and $\beta$ are integration constants. On the other hand,
the KK modes satisfy the following equation;
\begin{equation}
\frac{d^2 \rho^{(n)}}{dw^2}+\frac{d\log \Omega(w)}{dw}
\frac{d\rho^{(n)}}{dw}+m_n^2 \rho^{(n)}=0.
\label{kkeq}
\end{equation}
We will come back to this equation later.

Next we shall consider the effective field theory 
in terms of the zero mode on the $AdS_4$
brane. For that purpose we plug eq.(\ref{zmsol}) into the
original action (\ref{action}). The result is 
\begin{eqnarray}
S_{GF}^{(0)} &=& -\frac{1}{4}\int dx^4\int_{-w_L}^{w_L}dw
\sqrt{-G}G^{MN}G^{RS}F_{MR}^{(0)}F_{NS}^{(0)} \nonumber \\
&=& -\frac{1}{4}\int dx^4
\sqrt{-\bar{g}}\bar{g}^{\mu \nu}\bar{g}^{\lambda \sigma}
f_{\mu \lambda}^{(0)}f_{\nu \sigma}^{(0)}
\int_{-w_L}^{w_L}dw\Omega(w)(\rho^{(0)}(w))^2 \nonumber \\
& &-\frac{1}{4}\int dx^4 \sqrt{-\bar{g}}\bar{g}^{\mu \nu}
a_{\mu}^{(0)}a_{\nu}^{(0)}\int_{-w_L}^{w_L}dw
2\Omega(w)(\partial_w \rho^{(0)}(w))^2.
\label{zmaction}
\end{eqnarray}
This is the action of the massive gauge field in four dimensions.
We need to evaluate the $w$-integrations 
appearing in eq.(\ref{zmaction}) such as 
\begin{eqnarray}
I_1 &\equiv& \int_{-w_L}^{w_L}dw\Omega(w)(\rho^{(0)}(w))^2, \nonumber \\
I_2 &\equiv& \int_{-w_L}^{w_L}dw2\Omega(w)(\partial_w \rho^{(0)}(w))^2.
\label{integrals}
\end{eqnarray}
At first we compute the integration $I_2$ which is simpler.
\begin{eqnarray}
I_2 &=& \int_{-w_L}^{w_L}dw2\Omega(w)(\partial_w \rho^{(0)}(w))^2 
\nonumber \\
    &=& \frac{4\alpha^2}{\tilde{k}^2}\cosh(kz_0)
        \bigl[\sin\tilde{k}(w_L-\theta)+\sin\tilde{k}\theta\bigr].
\label{I2}
\end{eqnarray}
Next we evaluate the integration $I_1$ which is more complicated.
\begin{eqnarray}
I_1 &=& \int_{-w_L}^{w_L}dw\Omega(w)(\rho^{(0)}(w))^2 \nonumber \\
    &=& 2\int_{0}^{w_L}dw\frac{\Bigl(\frac{\alpha}{\tilde{k}}
    \cosh(kz_0)\sin\tilde{k}(w-\theta)+\beta\Bigr)^2}
    {\cosh(kz_0)\cos\tilde{k}(w-\theta)} \nonumber \\
    &=& \log F(w_L, \theta) -\frac{2\alpha^2}{\tilde{k}^3}\cosh(kz_0)
        \bigl[\sin\tilde{k}(w_L-\theta)+\sin\tilde{k}\theta\bigr],
\label{I1}    
\end{eqnarray}
where 
\begin{equation}
F(w_L, \theta)=
\Biggl[\frac{(1+\sin\tilde{k}(w_L-\theta))(1+\sin\tilde{k}\theta)}
{(1-\sin\tilde{k}(w_L-\theta))(1-\sin\tilde{k}\theta)}\Biggr]^a 
\Biggl[\frac{\cos\tilde{k}\theta}{\cos\tilde{k}(w_L-\theta)}\Biggr]^b
\label{F}
\end{equation}
with
\begin{eqnarray}
a &=& \frac{\alpha^2}{\tilde{k}^3}\cosh(kz_0)
      + \frac{\beta^2}{\tilde{k}\cosh(kz_0)},\nonumber \\
b &=& \frac{4\alpha\beta}{\tilde{k}^2}.
\label{a&b}
\end{eqnarray}
Unlike the case of \cite{oda}, this integral $I_1$ does not
diverge as long as $L$, which is the distance between two branes
in the original coordinate, is finite. Actually when $L$ is 
infinite, the integral diverges. We might say that
$L$ plays a role of an infrared cutoff in the model.

If we redefine the brane gauge field $a_{\mu}^{(0)}(x)$ as
\begin{equation}
\sqrt{I_1} a_{\mu}^{(0)} \longrightarrow a_{\mu}^{(0)},
\label{redef}
\end{equation}
we can rewrite the action in terms of canonically normalized field
as follows;
\begin{equation}
S_{GF}^{(0)} 
= -\frac{1}{4}\int dx^4
\sqrt{-\bar{g}}\Bigl[\bar{g}^{\mu \nu}\bar{g}^{\lambda \sigma}
f_{\mu \lambda}^{(0)}f_{\nu \sigma}^{(0)}
+\frac{I_2}{I_1}\bar{g}^{\mu \nu}a_{\mu}^{(0)}a_{\nu}^{(0)}\Bigr].
\label{zmaction2}
\end{equation}
Therefore we can  regard $\frac{I_2}{I_1}$ as the squared mass of 
the zero mode of the gauge field in four dimensions, which must be 
very tiny so as not to violate current experimental bounds.
For simplicity, let us consider the case of symmetric configuration, 
i.e., $w_L = 2\theta$. In this case, the mass of the gauge
field zero mode on the brane is of the form
\begin{equation}
M_{ZM}^2 = \frac{k^2}{\frac{1}{4\tanh(kz_0)}\Bigl(\cosh^2(kz_0)
+(\frac{\beta k}{\alpha})^2\frac{1}{\cosh^2(kz_0)}\Bigr)
\log\Bigl[\frac{1+\tanh(kz_0)}{1-\tanh(kz_0)}\Bigr]
-\frac{\cosh^2(kz_0)}{2}}.
\label{gaugemass}
\end{equation}
where we have used the relation $\tan(\frac{\tilde{k}\theta}{2})
= \tanh(\frac{kz_0}{2})$.

According to the Particle Data Group\cite{pdg}, 
present experimental bound on the photon mass is,  
\begin{center}
$M_{exp.} < 2~\times~10^{-16}~$(eV).~~~~(LAKES98)
\end{center}
Let us assume that $(\frac{\beta k}{\alpha})^2 \approx {\cal O}(1)$.
At $x \equiv kz_0 >> 1$,
\begin{eqnarray}
M_{ZM}^2(x) &\approx& \frac{k^2}{\frac{1}{4(1-2e^{-2x})}\frac{e^{2x}}{4}
\log\frac{1+(1-2e^{-2x})}{1-(1-2e^{-2x})}-\frac{e^{2x}}{8}} \nonumber \\
&=& e^{-2x}k^2\Biggl(\frac{x-1+2e^{-2x}}{8(1-2e^{-2x})}
+\frac{\log(1-e^{-2x})}{16(1-2e^{-2x})} \Biggr)^{-1} \nonumber \\
&\approx& \frac{8e^{-2x}}{x}k^2.
\label{approx}
\end{eqnarray}
Comparing the above experimental bound with 
eq.(\ref{approx}), we find that
we can make the mass of the gauge field small enough {\it if} 
we choose large enough $kz_0$, i.e. $kz_0 > 100$. This condition
gives a weaker constraint to the model than that coming from
present experimental and observational bounds on the four dimensional 
effective cosmological constant discussed in \cite{kogan}.

Finally let us consider the KK modes which satisfy eq.(\ref{kkeq}).
The general solution is given in terms of hypergeometric functions;
\begin{eqnarray}
\rho^{(n)}(w)&=&c_1 F\Bigl(\alpha_n, \beta_n, \frac{1}{2};
\sin^2 (\tilde{k}(|w|-\theta))\Bigr) \nonumber \\
& & \qquad \qquad +c_2 |\sin\tilde{k}(|w|-\theta)|
F\Bigl(\alpha_n +\frac{1}{2}, \beta_n +\frac{1}{2}, \frac{3}{2};
\sin^2 (\tilde{k}(|w|-\theta))\Bigr).
\label{hyper}
\end{eqnarray}
Here $c_1$ and $c_2$ are integration constants and
\begin{eqnarray}
\alpha_n &=& -\frac{1}{4}
+\frac{1}{2}\sqrt{\biggr(\frac{m_n}{\tilde{k}}\biggr)^2+\frac{1}{4}}, 
\nonumber \\
\beta_n &=& -\frac{1}{4}
-\frac{1}{2}\sqrt{\biggl(\frac{m_n}{\tilde{k}}\biggr)^2+\frac{1}{4}}.
\label{hyper2}
\end{eqnarray}
In the symmetric configuration, we have only to consider
even and odd functions with respect to the minimum of the
warp factor. In the case of the odd functions, we have $c_1=0$
while the even functions, we have $c_2=0$.


In summary, in this paper we investigated behavior of the bulk
gauge field in a recently proposed bigravity model. We found that
the zero mode of the gauge field is not constant but depends on
the fifth dimensional coordinate. The zero mode then
becomes massive on a brane and we require that the mass be 
tiny so as not to contradict current experimental bounds on 
the photon mass.  
This requirement leads us to $kz_0 > 100$. We also discussed the 
KK modes.
It will be interesting to study the bulk standard model where scalar
field(Higgs) and fermion fields(quarks and leptons) as well as gauge 
fields are in the bulk\cite{motoi}. 
We are also interested in the cosmology of this model.

\vspace{2cm}

\begin{center}
{\large{\bf Acknowledgments}}
\end{center}
We would like to thank Andreas Karch and Manolo Perez-Victoria
for useful conversations and comments. 
We also appreciate Zachary Guralnik for careful reading of this manuscript. 
This work was supported in part by a Grant-in-Aid for Scientific 
Research from Ministry of Education, Science, Sports and Culture 
of Japan (No.~3666).

\end{document}